\documentclass[aps, prl, twocolumn, groupedaddress,
  superscriptaddress, shortbibliography]{revtex4-1}

\usepackage{graphicx} 
\usepackage{color} 
\usepackage{amsmath}
\usepackage{amsfonts} 
\usepackage{amssymb} 
\usepackage{dcolumn} 
\usepackage[justification=centerlast]{caption}
\usepackage[draft]{todonotes} 
\usepackage{xr}
\usepackage{hyperref}
\hypersetup{colorlinks=true, urlcolor=blue,
  linkcolor=blue, citecolor=blue}

\newcommand{\vect}[1]{\mathbf{#1}}
\newcommand{\mychi}{\raisebox{0pt}[1ex][1ex]{$\chi$}}
\newcommand*\dif{\mathop{}\!\mathrm{d}}

\begin{document}
\title{Structural analysis of high-dimensional basins of attraction}
\author{Stefano Martiniani} 
\email{sm958@cam.ac.uk}
\affiliation{Department of Chemistry, University of Cambridge,
  Lensfield Road, Cambridge, CB2 1EW, UK}
\author{K. Julian Schrenk}
\affiliation{Department of Chemistry, University of Cambridge,
  Lensfield Road, Cambridge, CB2 1EW, UK}
\author{Jacob D. Stevenson}
\affiliation{Microsoft Research Ltd, 21 Station Road, Cambridge, CB1
  2FB, UK}
\affiliation{Department of Chemistry, University of Cambridge,
  Lensfield Road, Cambridge, CB2 1EW, UK}
\author{David J. Wales}
\affiliation{Department of Chemistry, University of Cambridge,
  Lensfield Road, Cambridge, CB2 1EW, UK}
\author{Daan Frenkel}
\affiliation{Department of Chemistry, University of Cambridge,
  Lensfield Road, Cambridge, CB2 1EW, UK}

\begin{abstract}
We propose an efficient Monte Carlo method for the computation of the
volumes of high-dimensional bodies with arbitrary shape. We start with
a region of known volume within the interior of the manifold and then
use the multi-state Bennett acceptance-ratio method to compute the
dimensionless free-energy difference between a series of equilibrium
simulations performed within this object. The method produces results
that are in excellent agreement with thermodynamic integration, as
well as a direct estimate of the associated statistical
uncertainties. The histogram method also allows us to directly obtain
an estimate of the interior radial probability density profile, thus
yielding useful insight into the structural properties of such a high
dimensional body. We illustrate the method by analysing the effect of
structural disorder on the basins of attraction of mechanically stable
packings of soft repulsive spheres.
\end{abstract}
\maketitle
\section{Introduction}
In science we often face, and occasionally confront, the following
question: ``Can we estimate the {\em a priori} probability of
observing a system in a very unlikely state?''  An example is: ``How
likely is a given disordered sphere packing?'', not to mention
questions such as ``How likely is life, or the existence of a universe
like ours?'' within the context of dynamical systems and of the
multiverse. In a number of cases, where the states
correspond to extrema in a high dimensional function, this question
can be narrowed down to: ``How large is the `basin of attraction' of a
given state?''.  In such cases, estimating the probability of
observing a particular state is equivalent to computing the volume of
the (high-dimensional) basin of attraction of this state. That
simplifies the problem, but not by much~\cite{ball1997elementary,
  simonovits2003compute}: analytical approaches are typically limited
to highly symmetric (often convex) volumes, whilst `brute force'
numerical techniques can deal with more complex shapes, but only in
low-dimensional cases.  Computing the volume of an arbitrary,
high-dimensional body is extremely challenging. For instance, it can
be proved that the exact computation of the volume of a convex
polytope is a NP-hard problem \cite{dyer1988complexity,
  khachiyan1989problem, khachiyan1988complexity} and, of course, the
problem does not get any easier in the non-convex case.

Yet, the importance of such computations is apparent: the volume of
the basin of attraction for the extrema of a generic energy landscape,
be that of biological molecules \cite{miller1999energy}, an artificial
neural network \cite{sagun2014explorations, ballard2016energy,
  ballard2016landscape}, a dynamical system \cite{wiley2006size,
  menck2013basin}, or even of a ``string theory landscape'' (where the
minima corresponds to different de Sitter vacua
\cite{frazer2011exploring, greene2013tumbling}), is essential for
understanding the systems' behavior.

In high dimensions, simple quadrature and brute-force sampling
fail~\cite{Bishop09} and other methods are needed.  In statistical
mechanics, the problem is equivalent to the calculation of the
partition function (or, equivalently, the free energy) of a system,
and several techniques have been developed to tackle this problem (see
e.g~\cite{Frenkel02}). The earliest class of techniques to compute
partition functions is based on thermodynamic integration (TI)
\cite{Kirkwood35,gelman1998simulating,Frenkel02}, which is based on
the idea that a transformation of the Hamiltonian of the system can
transform an unknown partition function into one that is known
analytically.  More recent techniques include histogram-based methods
(Wang-Landau \cite{PhysRevLett.86.2050}, parametric and non parametric
weighted histogram analysis method (WHAM) \cite{Habeck12}) or Nested
Sampling \cite{skilling2004nested,Martiniani14}.  In essence, all
these techniques reduce the computation of the partition function to
the numerical evaluation of a one-dimensional integral.

Among the above methods Nested Sampling and Wang Landau are Monte
Carlo algorithms in their own right, that produce the (binned) density
of states as a by-product. On the other hand, TI can be identified as
a particular Umbrella Sampling scheme~\cite{Frenkel02}, that outputs
multiple sets of equilibrium states that can be analysed, either by
numerical quadrature (e.g. see the Einstein crystal
method~\cite{Frenkel84}), or by WHAM and multi-state Bennet acceptance
ratio method (MBAR). All the above methods can be used to compute
high-dimensional volumes.  However, the choice of the MBAR method
\cite{Shirts2008} is an optimal one.  Not only is MBAR non-parametric
(no binning is required) and has the lowest known variance reweighting
estimator for free energy calculations, but it also eliminates the
need for explicit numerical integration of the density of states, thus
reducing to a minimum the number of systematic biases.

One reason why brute force methods are not suited to estimate the
volumes of high-dimensional bodies, is that for such bodies the volume
of the largest inscribed hypersphere, quickly becomes negligible to
the volume of the smallest circumscribed hypersphere -- and most of
the volume of the circumscribed hypersphere is empty.  Hence, using a
Monte Carlo `rejection method' to compute the volume of the non-convex
body as the fraction of volume contained in a
hypersphere~\cite{Sheng2007,Ashwin12}, does not yield accurate
results: the largest contribution should come from points that are
barely sampled, if at all.

In this Letter we show that MBAR can be used, not only to arrive at an
accurate estimate of a high-dimensional, non-convex volume, but that
it also can be used to probe the spatial distribution of this volume.

\section{Computing High-Dimensional Volumes}

Our aim is then to measure the volume of a $n-\text{dimensional}$
connected compact manifold $\Omega \subseteq \mathbb{R}^n$ with
boundaries. We require this body to be ``well guaranteed'', i.e. it
has both an inscribed and a circumscribed hypersphere
\cite{simonovits2003compute}.  To explore different parts of the
non-convex volume, we use a spherically symmetric bias that either
favors the sampling of points towards the center, or towards the
periphery.  We start by performing a series of $K+1$ random walks
under different applied bias potentials, similarly to the
Einstein-crystal method~\cite{Frenkel84}. We refer to each of the
walkers as a ``replica'' $R_i$. Unlike TI, where biasing is always
`attractive' (i.e. it favors larger confinement), in MBAR we are free
to choose both attractive and repulsive bias potentials (see SM for
details of our implementation). Additionally MBAR uses the full
posterior distribution (hence all moments) rather than just the
average log-likelihood computed over the posterior, as for TI. The
present method directly yields an estimate for the statistical
uncertainty in the results that depends on the full distributions and
is sensitive to their degree of overlap, thus making the method more
robust to under-sampling. In contrast, TI would require an expensive
resampling numerical procedure to achieve the same objective.

The Markov Chain Monte Carlo (MCMC) random walk of replica $i \in
[0,K]$ will generate samples with unnormalised probability density
$q_i(\vect{x})$, which for a standard Metropolis Monte Carlo walk is
\begin{equation} 
q_i(\vect{x}) \equiv e^{-\beta_iU_i(\vect{x})}
\end{equation}
with biasing potential $U_i(\vect{x})$ and inverse temperature
$\beta_i$; from now on we assume $\beta_i = 1$ for all walkers $R_i$,
without loss of generality. The normalised probability density is then
\begin{equation}
p_i(\vect{x}) = Z_i^{-1}q_i(\vect{x})
\end{equation}
with normalisation constant
\begin{equation}
\label{partition_function}
Z_i = \int_{\mathbb{R}^n} q_i(\vect{x}) \dif\vect{x}.
\end{equation}
We require that the bias potential $U_i(\vect{x})$ can be factorised as
\begin{equation}
U_i(\vect{x}) = \mathbf{\mychi}_{\Omega}(\vect{x})u_i(\vect{x})
\end{equation}
where $u_i$ is the reduced potential function and
$\mathbf{\mychi}_{\Omega}(\vect{x})$ is the ``oracle''
\cite{simonovits2003compute}, such that for all choices of $u_i(\vect{x})$,
\begin{equation}
U_i(\vect{x})= \left\{ 
  \begin{array}{l l}
    u_i(\vect{x}) & \quad \text{if $\vect{x} \in \Omega$} \\ 
    \infty & \quad \text{if $\vect{x} \not\in \Omega$}
  \end{array} \right.\ 
\end{equation}
We thus have that the normalisation constant in
Eq.~(\ref{partition_function}) becomes an integral over the manifold
$\Omega$
\begin{equation}
\label{partition_function2}
Z_i
= \int_{\mathbb{R}^n} e^{-U_i(\vect{x})} \dif \vect{x}
= \int_{\Omega} e^{-u_i(\vect{x})} \dif \vect{x}.
\end{equation}

If replica $R_M$ is chosen to have bias $u_M=0$, by definition
Eq.~(\ref{partition_function2}) becomes the volume $V_{\Omega}$. Hence
if we can compute the partition function for the reduced potential
function $u_M=0$, we can compute the volume $V_{\Omega}$.

The MBAR method \cite{Shirts2008} is a binless and statistically
optimal estimator to compute the difference in dimensionless free
energy for multiple sets of equilibrium states (trajectories) $\{
\vect{x} \}_i$ obtained using different biasing potentials
$u_i(\vect{x})$. The difference in dimensionless free energy is
defined as
\begin{equation}
\Delta \hat{f}_{ij} \equiv \hat{f}_{j} - \hat{f}_{i} = -\ln \left( \frac{Z_j}{Z_i}\right) 
\end{equation}
which can be computed by solving a set of self-consistent equations as
described in Ref. \cite{Shirts2008}. Note that only the differences of
the dimensionless free energies are meaningful as the absolute values
$\hat{f}_{i}$ are determined up to an additive constant and that the
``hat'' indicates MBAR estimates for the dimensionless free energies, to be
distinguished from the exact (reference) values.

Let us define the volume
$V_{\omega}=\pi^{n/2}r_{\omega}^{n}/\Gamma(n/2+1)$ of a $n$-ball
$\omega \subseteq \Omega$ with radius $r_{\omega}$ centred on
$\vect{x}_0$ and absolute dimensionless free energy $f_{\omega}=-\ln
V_{\omega}$. For instance, when the volume of a basin of attraction in
a potential energy landscape is to be measured, $\vect{x}_0$ is chosen
to be the minimum energy configuration and $\omega \subseteq \Omega$
the largest $n$-ball centred at $\vect{x}_0$ that fits in $\Omega$. We
also define $\{\vect{x}\}_{i}$ to be the set of states sampled with
biasing potential $u_i$ and $\{\vect{x}\}_{\omega} = \cup_{i=0}^K
\{\vect{x} : |\vect{x}-\vect{x}_0| \leq r_{\omega}\}_i$ to be the set
of states re-sampled within $\omega$ with reduced potential
\begin{equation}
u_\omega(\vect{x})= \left\{
  \begin{array}{l l}
    0 & \quad \text{if $|\vect{x}-\vect{x}_0| \leq r_{\omega}$} \\
    \infty & \quad \text{if $|\vect{x}-\vect{x}_0| > r_{\omega}$}
  \end{array} \right.\
\label{eq:resampling_potential} 
\end{equation}
In other words we augment the set of states with the additional
reduced potential $u_{\omega}$. Note that MBAR can compute free energy
differences and uncertainties between sets of states not sampled
(\emph{viz.} with a different reduced potential function) without any
additional iterative solution of the self-consistent estimating
equations, see Ref.~\cite{Shirts2008} for details.

Computing the free energy difference between the sets of equilibrium
states $\{\vect{x}\}_{\omega}$ and $\{\vect{x}\}_{M}$, chosen to have
reduced potentials $u_M=0$ and $u_{\omega}$, we find that the absolute
free energy for the unbiased set of states $\{\vect{x}\}_{M}$ is
\begin{equation}
f_{M} = f_{\omega} + (\hat{f}_{M} - \hat{f}_{\omega})  
\label{eq:free_energy_estimate}
\end{equation}
where the free energy difference $\hat{f}_{M} - \hat{f}_{\omega}$ is
obtained by MBAR with associated uncertainty
$\delta\Delta\hat{f}_{M\omega}$. The volume of the manifold is then
just $V_{\Omega} = \exp(-f_M)$ with uncertainty $\delta
V_{\Omega}=V_{\Omega}\delta\Delta\hat{f}_{M\omega}$. Note that the set
of biasing potentials $u_i$ must be chosen so that there is sufficient
overlap between each neighbouring pair of $p_i(\vect{x})$. For
instance for the harmonic bias $u_i = k_i|\vect{x}-\vect{x}_0|^2/2$ we
must choose a set of coupling constants $k_i$ so that all neighbouring
replicas have a sufficient probability density overlap.

Under an appropriate choice of biasing potential the present method
may yield information such as the radial posterior probability density
function, as an easy to compute by-product, details are discussed in
the SM.

\section{Basins of attraction in high dimensions}

We define a basin of attraction as the set of all points that lead to
a particular minimum energy configuration by a path of steepest
descent on a potential energy surface (PES). Exploring a basin of
attraction is computationally expensive because each call to the
oracle function $\mathbf{\mychi}_{\Omega}(\vect{x})$ requires a full
energy minimisation and equilibrating a MCMC on a high dimensional
support is difficult \cite{Xu11, Asenjo13, Asenjo14,
  Martiniani16}. For this reason little is known about the geometry of
these bodies \cite{Xu10, Wang12, Asenjo13, Martiniani16}.

Ashwin et al. \cite{Ashwin12}, defined the basin of attraction as the
collection of initial zero-density configurations that evolve to a
given jammed packing of soft repulsive disks via a compressive quench.
On the basis of `brute-force' calculations on low-dimensional systems,
Ashwin et al. suggested that basins of attraction tend to be
``branched and threadlike'' away from a spherical core
region. However, the approach of ref.~\cite{Ashwin12} breaks down for
higher dimensional systems for which most of the volume of the basin
is concentrated at distances from the `minimum' where the overwhelming
majority of points do not belong to the basin. The method that we
present here allows us to explore precisely those very rarified
regions where most of the `mass' of a basin is concentrated.

In general the representation of \emph{all} high dimensional
\emph{convex} bodies should have a hyperbolic form such as the one
proposed in the illustration by Ashwin et al. due to the exponential
decay in volume of parallel hypersections (slices) away from the
median (or equator) \cite{milman1998surprising}. This holds true even
for the simplest convex bodies, such as the hypercube, and the
underlying geometry need not be ``complicated'', as one would guess at
first from the two-dimensional representation. For the simplest cases
of the unit $d$-sphere and the unit $d$-cube it can be shown that most
of the volume is contained within $\mathcal{O}(1/d)$ of the boundary
and that at the same time the volume is contained in a slab
$\mathcal{O}(1/\sqrt{d})$ and $\mathcal{O}(1)$ from the equator,
irrespective of the choice of north pole, respectively
\cite{ball1997elementary, Guruswami2012}.  Hence, there is virtually
no interior volume. Such phenomena of concentration of measure are
ubiquitous in high dimensional geometry and are closely related to the
law of large numbers \cite{Guruswami2012}.

As we will show, the results presented by Ashwin et al. are, within
the resolution available to their method, qualitatively consistent
with those for a simple (unit) hypercube.

\begin{figure}
    \includegraphics[width=\columnwidth]{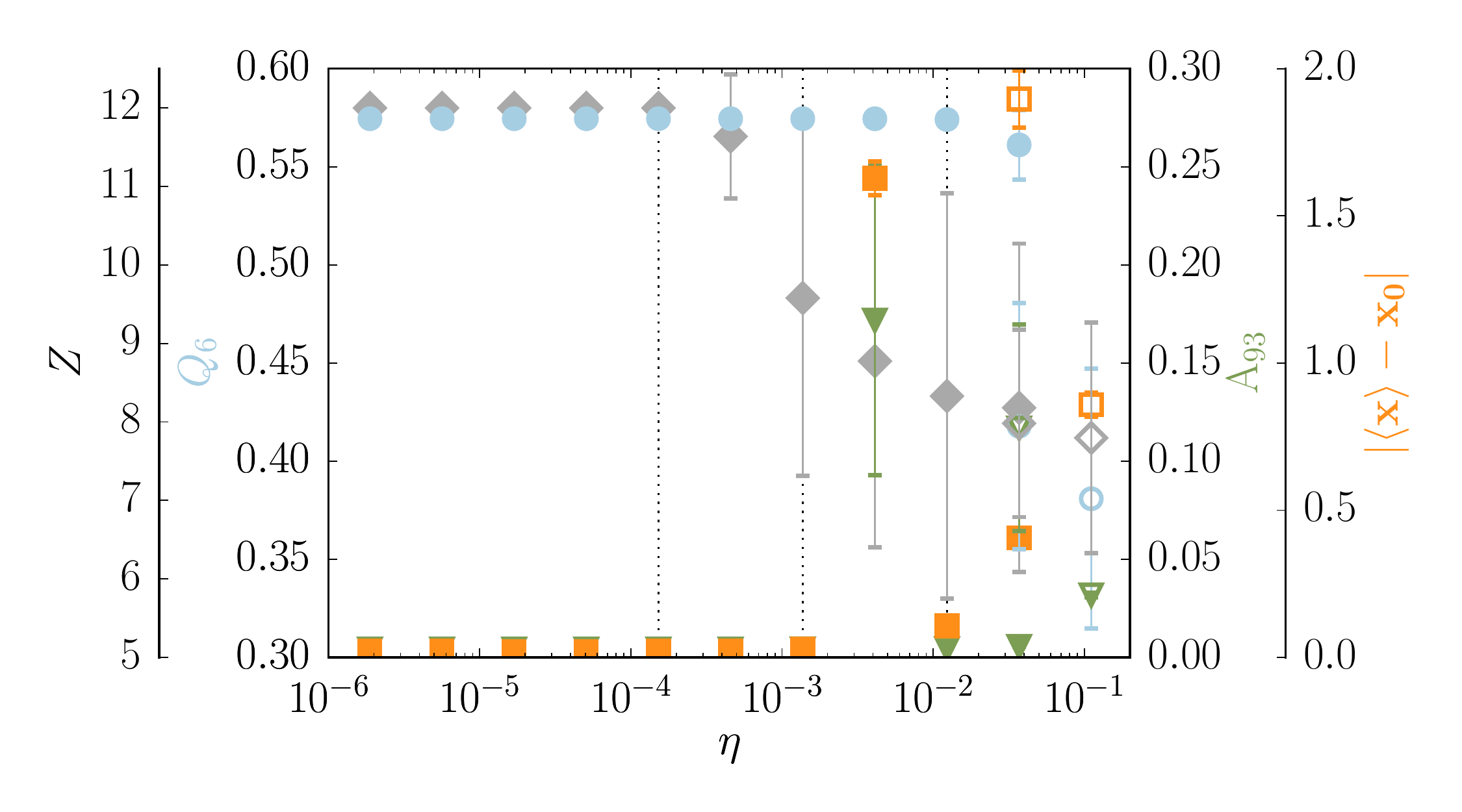}
    \caption{\label{fig::q6} Structural disorder as a function of
      polydispersity $\eta$ is quantified by the average coordination
      number $Z$ (grey diamonds) and the $Q_6$ bond orientational
      order parameter (blue circles); error bars correspond to one
      standard deviation of the distribution of values per
      particle. Basin shape is characterized by the asphericity factor
      $A_d$ (green triangles) and the mean distance of the centre of
      mass from the minmum (orange squares); error bars correspond to
      the standard error. Filled and empty markers correspond to
      packings obtained starting from an fcc and a disordered
      arragement respectively. Dotted lines show the $\eta$ after
      which, in order, $Z$, $A_d$ and $Q_6$ change from the fcc
      value.}
\end{figure}

\begin{figure}
    \includegraphics[width=\columnwidth]{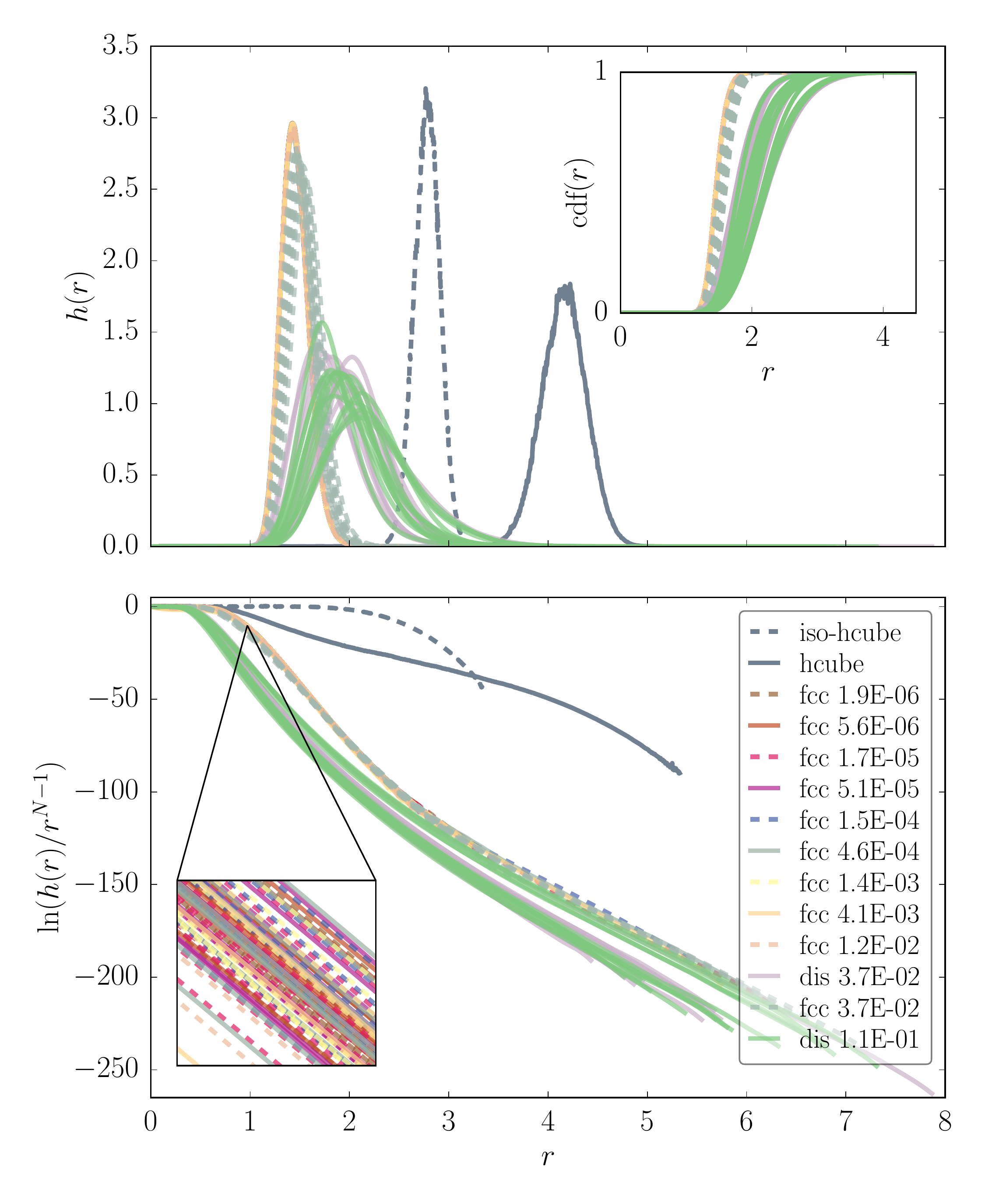}
    \caption{\label{fig::dos}Top plot shows the measured basin radial
      probability density function $h(r)$ (DOS) for packings at
      different polydispersities. The solid and dashed blue curves
      correspond to the DOS of a $93$D hypercube, measured from the
      centre of mass (`iso-cube') and from a point in one of the
      corners. The top inset shows the cumulative distribution
      function for $h(r)$. The bottom panel shows the logarithm of the
      ratio of the DOS of the basin and of a $93$D hyperball. The
      bottom inset shows the set of barely distinguishable overlapping
      curves measured for low polydispersities. Top and bottom plots
      share the x-axis.}
\end{figure}

\subsection{Effect of structural disorder on the basins of attraction
  of jammed sphere packings}

We characterise the basins of attraction for a number of 32 hard-core
plus soft-shell three-dimensional sphere packings, analogous to the
ones described in Ref.~\cite{Martiniani16}. The soft shell
interactions are short ranged and purely repulsive, the full
functional form of the potential and further technical details are
reported in the SM. We systematically introduce structural disorder by
preparing packings with (geometrically) increasing particle size
polydispersity $\eta$, i.e. the (positive) radii are sampled from a
normal distribution $\mathcal{N}(1,\eta)$. For each $\eta$ we prepare
$\sim$10 packings at a soft packing fraction $\phi=0.74148$ with a
soft to hard-sphere radius ratio of $r_{\mathrm{SS}}/r_{\mathrm{HS}} =
1.12$. The particles are placed initially in a fcc arrangement
$\vect{x}_{\text{fcc}}$ and then relaxed via an energy minimisation to
a mechanically stable state $\vect{x}_{0}$. Thus, for the lowest
polydispersities the packings remain in a perfect fcc structure and
with increasing $\eta$ they progressively move away into a disordered
glassy state. For the largest polydispersity, for which hard-core
overlaps do not allow an initial fcc arrangement, we sample a series
of completely random initial states followed by an energy
minimisation. Note that even for $\eta \approx 0$, due to the high
packing fraction, starting from a completely random set of
coordinates, an energy minimisation does not lead to the fcc crystal
but rather to the closest glassy state (inherent structure). We are
interested in the effect of structural disorder on the shape of the
basin of attraction for the soft sphere packings.

We determine the amount of structural disorder in the packing by
computing the $Q_6$ bond orientational order parameter
\cite{steinhardt1983bond} and the average number of contacts per
particle $Z$, shown in Fig.~\ref{fig::q6}.  As the polydispersity of
the system is increased, the coordination number $Z$ decays
monotonically from the close-packed value of $12$ to a value
$Z_{\mathrm{fcc}}>Z>Z_{\mathrm{iso}}$, where $Z_{\mathrm{iso}}=6$ is
the average contact number at iso-staticity for a three-dimensional
packing of frictionless spheres \cite{OHern03}. The $Q_6$ order
parameter, computed using a solid-angle based nearest-neighbor
definition \cite{vanMeel12}, decays from its fcc value well after the
contact number has dropped below the close-packed value of $12$.

We start characterising the shape of the high dimensional basins of
attraction associated with these packings by performing an
unconstrained random walk within the basin and performing principal
component analysis (PCA) on the trajectory thus obtained
\cite{Bishop09}. PCA yields a set of eigenvectors that span the
$d$-dimensional configurational space with associated eigenvalues
$\lambda_1,\dots,\lambda_{d}$. If the basin posses $d$-dimensional
spherical symmetry then all the eigenvalues are expected to be
equal. A measure of the shape of a random walk is then the asphericity
factor \cite{rudnick1987shapes}
\begin{equation}
A_d = \frac{\sum_{i>j}(\lambda_i-\lambda_j)^2}{(d-1)\left(\sum_{i=1}^d
  \lambda_i\right)^2},
\end{equation} 
that has a value of $0$ for a spherically symmetric random walk and of
$1$ for a walk that extends only in one dimension. Furthermore, we
compute the distance of the centre of mass (\emph{CoM}) position from
the minimum energy configuration for the random walk, $|\langle
\vect{x} \rangle - \vect{x}_0|$. This quantity reveals whether the
basin is isotropic around the minimum or not. Both quantities,
averaged over all packings, are plotted as a function of
polydispersity in Fig.~\ref{fig::q6} along with the structural order
parameters. Interestingly, we observe that for low $\eta$ the basins
are, on average, spherically symmetric and isotropic around the
minimum. With the onset of structural disorder we observe a marginal
increase in asphericity and in the \emph{CoM} distance from the
minimum. In order to observe a significant change however, we need to
go to the fully disordered packings at higher polydispersity.  With
increasing polydispersity, we observe significant changes in the
structural order parameters and in the asphericity factor $A_d$ and
\emph{CoM} distance from the minimum.

The implementation details of the MBAR method that we have used are
discussed in the SM. Using this method to compute the volume of the
basins of attraction, we find excellent agreement with thermodynamic
integration, see Fig.~S2. As a natural by-product of the computation
we are able to compute the radial probability density function (DOS),
shown in Fig.~\ref{fig::dos} together with the logarithm of the ratio
between the measured DOS, and that of a $d$-hypersphere. The log-ratio
curves clearly show that all basins have a well-defined hyperspherical
core region, where the curves are flat around $0$, followed by a
series of exponential decays at larger distances from the minimum. For
$\eta < 10^{-4}$ the curves are mostly indistinguishable from one
another with most of the probability mass concentrated between $ 1< r
< 3$, as it can be seen from the inset showing the corresponding
cumulative distribution function (CDF). For higher polydispersity, the
DOS curves have ever longer tails, as it is also shown by the
systematic shift in the CDF.

Importantly, the curves show that a `rejection' method to measure the
basin volume will fail. In this method, the volume of the basin is
determined by integrating the fraction of points on a hyper-shell with
radius $r$ that fall inside the basin. That fraction is the function
shown in the bottom panel of Fig.~\ref{fig::dos}. The most important
contribution to the integral would come from the range of $r$ values
where $h(r)$ (top panel of Fig.~\ref{fig::dos}) has a significant
value. As can be seen from the figure, for disordered systems this
happens for values of $r$ where the fraction of hyper-sphere points
within the basin is extremely small, in the example shown
$\mathcal{O}(10^{-30})$. Hence, the dominant part of the integral
would come from parts that are never sampled.

To interpret our results for the DOS curves, it is useful to compare
with the corresponding result for a unit hypercube (see
Fig.~\ref{fig::dos}). In one instance we do so by placing the `origin'
of the hypercube at its \emph{CoM}, and in another by placing the
origin on one of the $2^d$ corners of the hypercube, to generate a DOS
of a system with a very anisometric density distribution.  Not
surprisingly, moving the origin of the system from the center to the
corner of a hypercube has a dramatic effect on the shape of the DOS,
which is now much more similar to the curves for large $\eta$, with
similar characteristic changes of slope observed for the
basins. Again, this agrees with the observation that the \emph{CoM}
distance increases with increasing structural disorder. The effect of
the basin asphericity, as measured by the asphericity factor $A_d$ is
difficult to infer from the DOS alone.

We thus observe that the structural isotropy and high degree of
rotational symmetry in the crystal, as indicated by the $Q_6$
parameter, is reflected in the isotropy and spherical symmetry of the
basin around the minimum, even for relatively large polydispersities
when the average contact number has already dropped considerably from
the close-packed value. Similarly, the structural disorder at larger
$\eta$ is reflected in the anisotropy and asphericity of the
basin. Hence, changes in the basin structure, as indicated by the
asphericity factor, the $CoM$ and the density profile, occur before
any observable changes occur in $Q_6$ and after the average contact
number ($Z \lesssim 9$) has fallen well below the close-packed value
of $12$.

\begin{acknowledgments}
S.M. acknowledges financial support by the Gates Cambridge
Scholarship.  K.J.S. acknowledges support by the Swiss National
Science Foundation under Grant No. P2EZP2-152188 and
No. P300P2-161078.  J.D.S. acknowledges support by Marie Curie Grant
275544.  D.F. and D.J.W. acknowledge support by EPSRC Programme Grant
EP/I001352/1, by EPSRC grant EP/I000844/1 (D.F.) and ERC Advanced
Grant RG59508 (D.J.W.).
\end{acknowledgments}
\bibliography{mbar_bv_method}

\begin{thebibliography}{37}%
\makeatletter
\providecommand \@ifxundefined [1]{%
 \@ifx{#1\undefined}
}%
\providecommand \@ifnum [1]{%
 \ifnum #1\expandafter \@firstoftwo
 \else \expandafter \@secondoftwo
 \fi
}%
\providecommand \@ifx [1]{%
 \ifx #1\expandafter \@firstoftwo
 \else \expandafter \@secondoftwo
 \fi
}%
\providecommand \natexlab [1]{#1}%
\providecommand \enquote  [1]{``#1''}%
\providecommand \bibnamefont  [1]{#1}%
\providecommand \bibfnamefont [1]{#1}%
\providecommand \citenamefont [1]{#1}%
\providecommand \href@noop [0]{\@secondoftwo}%
\providecommand \href [0]{\begingroup \@sanitize@url \@href}%
\providecommand \@href[1]{\@@startlink{#1}\@@href}%
\providecommand \@@href[1]{\endgroup#1\@@endlink}%
\providecommand \@sanitize@url [0]{\catcode `\\12\catcode `\$12\catcode
  `\&12\catcode `\#12\catcode `\^12\catcode `\_12\catcode `\%12\relax}%
\providecommand \@@startlink[1]{}%
\providecommand \@@endlink[0]{}%
\providecommand \url  [0]{\begingroup\@sanitize@url \@url }%
\providecommand \@url [1]{\endgroup\@href {#1}{\urlprefix }}%
\providecommand \urlprefix  [0]{URL }%
\providecommand \Eprint [0]{\href }%
\providecommand \doibase [0]{http://dx.doi.org/}%
\providecommand \selectlanguage [0]{\@gobble}%
\providecommand \bibinfo  [0]{\@secondoftwo}%
\providecommand \bibfield  [0]{\@secondoftwo}%
\providecommand \translation [1]{[#1]}%
\providecommand \BibitemOpen [0]{}%
\providecommand \bibitemStop [0]{}%
\providecommand \bibitemNoStop [0]{.\EOS\space}%
\providecommand \EOS [0]{\spacefactor3000\relax}%
\providecommand \BibitemShut  [1]{\csname bibitem#1\endcsname}%
\let\auto@bib@innerbib\@empty
\bibitem [{\citenamefont {Ball}(1997)}]{ball1997elementary}%
  \BibitemOpen
  \bibfield  {author} {\bibinfo {author} {\bibfnamefont {K.}~\bibnamefont
  {Ball}},\ }\href@noop {} {\bibfield  {journal} {\bibinfo  {journal} {Flavors
  of geometry}\ }\textbf {\bibinfo {volume} {31}},\ \bibinfo {pages} {1}
  (\bibinfo {year} {1997})}\BibitemShut {NoStop}%
\bibitem [{\citenamefont {Simonovits}(2003)}]{simonovits2003compute}%
  \BibitemOpen
  \bibfield  {author} {\bibinfo {author} {\bibfnamefont {M.}~\bibnamefont
  {Simonovits}},\ }\href@noop {} {\bibfield  {journal} {\bibinfo  {journal}
  {Mathematical programming}\ }\textbf {\bibinfo {volume} {97}},\ \bibinfo
  {pages} {337} (\bibinfo {year} {2003})}\BibitemShut {NoStop}%
\bibitem [{\citenamefont {Dyer}\ and\ \citenamefont
  {Frieze}(1988)}]{dyer1988complexity}%
  \BibitemOpen
  \bibfield  {author} {\bibinfo {author} {\bibfnamefont {M.~E.}\ \bibnamefont
  {Dyer}}\ and\ \bibinfo {author} {\bibfnamefont {A.~M.}\ \bibnamefont
  {Frieze}},\ }\href {\doibase 10.1137/0217060} {\bibfield  {journal} {\bibinfo
   {journal} {SIAM Journal on Computing}\ }\textbf {\bibinfo {volume} {17}},\
  \bibinfo {pages} {967} (\bibinfo {year} {1988})}\BibitemShut {NoStop}%
\bibitem [{\citenamefont {Khachiyan}(1989)}]{khachiyan1989problem}%
  \BibitemOpen
  \bibfield  {author} {\bibinfo {author} {\bibfnamefont {L.~G.}\ \bibnamefont
  {Khachiyan}},\ }\href@noop {} {\bibfield  {journal} {\bibinfo  {journal}
  {Uspekhi Mat. Nauk}\ }\textbf {\bibinfo {volume} {44}},\ \bibinfo {pages}
  {199} (\bibinfo {year} {1989})}\BibitemShut {NoStop}%
\bibitem [{\citenamefont {Khachiyan}(1988)}]{khachiyan1988complexity}%
  \BibitemOpen
  \bibfield  {author} {\bibinfo {author} {\bibfnamefont {L.~G.}\ \bibnamefont
  {Khachiyan}},\ }\href@noop {} {\bibfield  {journal} {\bibinfo  {journal}
  {Izvestia Akad. Nauk SSSR, Engineering Cybernetics}\ }\textbf {\bibinfo
  {volume} {3}},\ \bibinfo {pages} {216} (\bibinfo {year} {1988})}\BibitemShut
  {NoStop}%
\bibitem [{\citenamefont {Miller}\ and\ \citenamefont
  {Wales}(1999)}]{miller1999energy}%
  \BibitemOpen
  \bibfield  {author} {\bibinfo {author} {\bibfnamefont {M.~A.}\ \bibnamefont
  {Miller}}\ and\ \bibinfo {author} {\bibfnamefont {D.~J.}\ \bibnamefont
  {Wales}},\ }\href {\doibase 10.1063/1.480011} {\bibfield  {journal} {\bibinfo
   {journal} {Journal of Chemical Physics}\ }\textbf {\bibinfo {volume}
  {111}},\ \bibinfo {pages} {6610} (\bibinfo {year} {1999})}\BibitemShut
  {NoStop}%
\bibitem [{\citenamefont {Sagun}\ \emph {et~al.}(2014)\citenamefont {Sagun},
  \citenamefont {Guney},\ and\ \citenamefont {LeCun}}]{sagun2014explorations}%
  \BibitemOpen
  \bibfield  {author} {\bibinfo {author} {\bibfnamefont {L.}~\bibnamefont
  {Sagun}}, \bibinfo {author} {\bibfnamefont {V.~U.}\ \bibnamefont {Guney}}, \
  and\ \bibinfo {author} {\bibfnamefont {Y.}~\bibnamefont {LeCun}},\
  }\href@noop {} {\bibfield  {journal} {\bibinfo  {journal} {arXiv:1412.6615}\
  } (\bibinfo {year} {2014})}\BibitemShut {NoStop}%
\bibitem [{\citenamefont {Ballard}\ \emph
  {et~al.}(2016{\natexlab{a}})\citenamefont {Ballard}, \citenamefont
  {Stevenson}, \citenamefont {Das},\ and\ \citenamefont
  {Wales}}]{ballard2016energy}%
  \BibitemOpen
  \bibfield  {author} {\bibinfo {author} {\bibfnamefont {A.}~\bibnamefont
  {Ballard}}, \bibinfo {author} {\bibfnamefont {J.~D.}\ \bibnamefont
  {Stevenson}}, \bibinfo {author} {\bibfnamefont {R.}~\bibnamefont {Das}}, \
  and\ \bibinfo {author} {\bibfnamefont {D.~J.}\ \bibnamefont {Wales}},\
  }\href@noop {} {\bibfield  {journal} {\bibinfo  {journal} {The Journal of
  Chemical Physics}\ ,\ \bibinfo {pages} {accepted}} (\bibinfo {year}
  {2016}{\natexlab{a}})}\BibitemShut {NoStop}%
\bibitem [{\citenamefont {Ballard}\ \emph
  {et~al.}(2016{\natexlab{b}})\citenamefont {Ballard}, \citenamefont
  {Stevenson},\ and\ \citenamefont {Wales}}]{ballard2016landscape}%
  \BibitemOpen
  \bibfield  {author} {\bibinfo {author} {\bibfnamefont {A.}~\bibnamefont
  {Ballard}}, \bibinfo {author} {\bibfnamefont {J.~D.}\ \bibnamefont
  {Stevenson}}, \ and\ \bibinfo {author} {\bibfnamefont {D.~J.}\ \bibnamefont
  {Wales}},\ }\href@noop {} {\bibfield  {journal} {\bibinfo  {journal} {in
  submission}\ } (\bibinfo {year} {2016}{\natexlab{b}})}\BibitemShut {NoStop}%
\bibitem [{\citenamefont {Wiley}\ \emph {et~al.}(2006)\citenamefont {Wiley},
  \citenamefont {Strogatz},\ and\ \citenamefont {Girvan}}]{wiley2006size}%
  \BibitemOpen
  \bibfield  {author} {\bibinfo {author} {\bibfnamefont {D.~A.}\ \bibnamefont
  {Wiley}}, \bibinfo {author} {\bibfnamefont {S.~H.}\ \bibnamefont {Strogatz}},
  \ and\ \bibinfo {author} {\bibfnamefont {M.}~\bibnamefont {Girvan}},\
  }\href@noop {} {\bibfield  {journal} {\bibinfo  {journal} {Chaos: An
  Interdisciplinary Journal of Nonlinear Science}\ }\textbf {\bibinfo {volume}
  {16}},\ \bibinfo {pages} {015103} (\bibinfo {year} {2006})}\BibitemShut
  {NoStop}%
\bibitem [{\citenamefont {Menck}\ \emph {et~al.}(2013)\citenamefont {Menck},
  \citenamefont {Heitzig}, \citenamefont {Marwan},\ and\ \citenamefont
  {Kurths}}]{menck2013basin}%
  \BibitemOpen
  \bibfield  {author} {\bibinfo {author} {\bibfnamefont {P.~J.}\ \bibnamefont
  {Menck}}, \bibinfo {author} {\bibfnamefont {J.}~\bibnamefont {Heitzig}},
  \bibinfo {author} {\bibfnamefont {N.}~\bibnamefont {Marwan}}, \ and\ \bibinfo
  {author} {\bibfnamefont {J.}~\bibnamefont {Kurths}},\ }\href@noop {}
  {\bibfield  {journal} {\bibinfo  {journal} {Nature Physics}\ }\textbf
  {\bibinfo {volume} {9}},\ \bibinfo {pages} {89} (\bibinfo {year}
  {2013})}\BibitemShut {NoStop}%
\bibitem [{\citenamefont {Frazer}\ and\ \citenamefont
  {Liddle}(2011)}]{frazer2011exploring}%
  \BibitemOpen
  \bibfield  {author} {\bibinfo {author} {\bibfnamefont {J.}~\bibnamefont
  {Frazer}}\ and\ \bibinfo {author} {\bibfnamefont {A.~R.}\ \bibnamefont
  {Liddle}},\ }\href@noop {} {\bibfield  {journal} {\bibinfo  {journal}
  {Journal of Cosmology and Astroparticle Physics}\ }\textbf {\bibinfo {volume}
  {2011}},\ \bibinfo {pages} {026} (\bibinfo {year} {2011})}\BibitemShut
  {NoStop}%
\bibitem [{\citenamefont {Greene}\ \emph {et~al.}(2013)\citenamefont {Greene},
  \citenamefont {Kagan}, \citenamefont {Masoumi}, \citenamefont {Mehta},
  \citenamefont {Weinberg},\ and\ \citenamefont {Xiao}}]{greene2013tumbling}%
  \BibitemOpen
  \bibfield  {author} {\bibinfo {author} {\bibfnamefont {B.}~\bibnamefont
  {Greene}}, \bibinfo {author} {\bibfnamefont {D.}~\bibnamefont {Kagan}},
  \bibinfo {author} {\bibfnamefont {A.}~\bibnamefont {Masoumi}}, \bibinfo
  {author} {\bibfnamefont {D.}~\bibnamefont {Mehta}}, \bibinfo {author}
  {\bibfnamefont {E.~J.}\ \bibnamefont {Weinberg}}, \ and\ \bibinfo {author}
  {\bibfnamefont {X.}~\bibnamefont {Xiao}},\ }\href@noop {} {\bibfield
  {journal} {\bibinfo  {journal} {Physical Review D}\ }\textbf {\bibinfo
  {volume} {88}},\ \bibinfo {pages} {026005} (\bibinfo {year}
  {2013})}\BibitemShut {NoStop}%
\bibitem [{\citenamefont {Bishop}(2009)}]{Bishop09}%
  \BibitemOpen
  \bibfield  {author} {\bibinfo {author} {\bibfnamefont {C.~M.}\ \bibnamefont
  {Bishop}},\ }\href@noop {} {\emph {\bibinfo {title} {Pattern recognition and
  machine learning}}}\ (\bibinfo  {publisher} {Springer},\ \bibinfo {address}
  {New York},\ \bibinfo {year} {2009})\BibitemShut {NoStop}%
\bibitem [{\citenamefont {Frenkel}\ and\ \citenamefont
  {Smit}(2002)}]{Frenkel02}%
  \BibitemOpen
  \bibfield  {author} {\bibinfo {author} {\bibfnamefont {D.}~\bibnamefont
  {Frenkel}}\ and\ \bibinfo {author} {\bibfnamefont {B.}~\bibnamefont {Smit}},\
  }\href@noop {} {\emph {\bibinfo {title} {Understanding molecular
  simulation}}}\ (\bibinfo  {publisher} {Academic Press},\ \bibinfo {address}
  {San Diego},\ \bibinfo {year} {2002})\BibitemShut {NoStop}%
\bibitem [{\citenamefont {Kirkwood}(1935)}]{Kirkwood35}%
  \BibitemOpen
  \bibfield  {author} {\bibinfo {author} {\bibfnamefont {J.~G.}\ \bibnamefont
  {Kirkwood}},\ }\href {\doibase 10.1063/1.1749657} {\bibfield  {journal}
  {\bibinfo  {journal} {Journal of Chemical Physics}\ }\textbf {\bibinfo
  {volume} {3}},\ \bibinfo {pages} {300} (\bibinfo {year} {1935})}\BibitemShut
  {NoStop}%
\bibitem [{\citenamefont {Gelman}\ and\ \citenamefont
  {Meng}(1998)}]{gelman1998simulating}%
  \BibitemOpen
  \bibfield  {author} {\bibinfo {author} {\bibfnamefont {A.}~\bibnamefont
  {Gelman}}\ and\ \bibinfo {author} {\bibfnamefont {X.-L.}\ \bibnamefont
  {Meng}},\ }\href {http://www.jstor.org/stable/2676756} {\bibfield  {journal}
  {\bibinfo  {journal} {Statistical Science}\ }\textbf {\bibinfo {volume}
  {13}},\ \bibinfo {pages} {163} (\bibinfo {year} {1998})}\BibitemShut
  {NoStop}%
\bibitem [{\citenamefont {Wang}\ and\ \citenamefont
  {Landau}(2001)}]{PhysRevLett.86.2050}%
  \BibitemOpen
  \bibfield  {author} {\bibinfo {author} {\bibfnamefont {F.}~\bibnamefont
  {Wang}}\ and\ \bibinfo {author} {\bibfnamefont {D.~P.}\ \bibnamefont
  {Landau}},\ }\href {\doibase 10.1103/PhysRevLett.86.2050} {\bibfield
  {journal} {\bibinfo  {journal} {Physical Review Letters}\ }\textbf {\bibinfo
  {volume} {86}},\ \bibinfo {pages} {2050} (\bibinfo {year}
  {2001})}\BibitemShut {NoStop}%
\bibitem [{\citenamefont {Habeck}(2012)}]{Habeck12}%
  \BibitemOpen
  \bibfield  {author} {\bibinfo {author} {\bibfnamefont {M.}~\bibnamefont
  {Habeck}},\ }in\ \href@noop {} {\emph {\bibinfo {booktitle} {International
  Conference on Artificial Intelligence and Statistics}}}\ (\bibinfo {year}
  {2012})\ pp.\ \bibinfo {pages} {486--494}\BibitemShut {NoStop}%
\bibitem [{\citenamefont {Skilling}(2004)}]{skilling2004nested}%
  \BibitemOpen
  \bibfield  {author} {\bibinfo {author} {\bibfnamefont {J.}~\bibnamefont
  {Skilling}},\ }\href {\doibase 10.1063/1.1835238} {\bibfield  {journal}
  {\bibinfo  {journal} {AIP Conf. Proc. Bayesian inference and maximum entropy
  methods in science and engineering}\ }\textbf {\bibinfo {volume} {735}},\
  \bibinfo {pages} {395} (\bibinfo {year} {2004})}\BibitemShut {NoStop}%
\bibitem [{\citenamefont {Martiniani}\ \emph {et~al.}(2014)\citenamefont
  {Martiniani}, \citenamefont {Stevenson}, \citenamefont {Wales},\ and\
  \citenamefont {Frenkel}}]{Martiniani14}%
  \BibitemOpen
  \bibfield  {author} {\bibinfo {author} {\bibfnamefont {S.}~\bibnamefont
  {Martiniani}}, \bibinfo {author} {\bibfnamefont {J.~D.}\ \bibnamefont
  {Stevenson}}, \bibinfo {author} {\bibfnamefont {D.~J.}\ \bibnamefont
  {Wales}}, \ and\ \bibinfo {author} {\bibfnamefont {D.}~\bibnamefont
  {Frenkel}},\ }\href {\doibase 10.1103/PhysRevX.4.031034} {\bibfield
  {journal} {\bibinfo  {journal} {Physical Review X}\ }\textbf {\bibinfo
  {volume} {4}},\ \bibinfo {pages} {031034} (\bibinfo {year}
  {2014})}\BibitemShut {NoStop}%
\bibitem [{\citenamefont {Frenkel}\ and\ \citenamefont
  {Ladd}(1984)}]{Frenkel84}%
  \BibitemOpen
  \bibfield  {author} {\bibinfo {author} {\bibfnamefont {D.}~\bibnamefont
  {Frenkel}}\ and\ \bibinfo {author} {\bibfnamefont {A.~J.~C.}\ \bibnamefont
  {Ladd}},\ }\href {\doibase 10.1063/1.448024} {\bibfield  {journal} {\bibinfo
  {journal} {Journal of Chemical Physics}\ }\textbf {\bibinfo {volume} {81}},\
  \bibinfo {pages} {3188} (\bibinfo {year} {1984})}\BibitemShut {NoStop}%
\bibitem [{\citenamefont {Shirts}\ and\ \citenamefont
  {Chodera}(2008)}]{Shirts2008}%
  \BibitemOpen
  \bibfield  {author} {\bibinfo {author} {\bibfnamefont {M.~R.}\ \bibnamefont
  {Shirts}}\ and\ \bibinfo {author} {\bibfnamefont {J.~D.}\ \bibnamefont
  {Chodera}},\ }\href {\doibase 10.1063/1.2978177} {\bibfield  {journal}
  {\bibinfo  {journal} {Journal of Chemical Physics}\ }\textbf {\bibinfo
  {volume} {129}},\ \bibinfo {eid} {124105} (\bibinfo {year}
  {2008})}\BibitemShut {NoStop}%
\bibitem [{\citenamefont {Liu}\ \emph {et~al.}(2007)\citenamefont {Liu},
  \citenamefont {Zhang},\ and\ \citenamefont {Zhu}}]{Sheng2007}%
  \BibitemOpen
  \bibfield  {author} {\bibinfo {author} {\bibfnamefont {S.}~\bibnamefont
  {Liu}}, \bibinfo {author} {\bibfnamefont {J.}~\bibnamefont {Zhang}}, \ and\
  \bibinfo {author} {\bibfnamefont {B.}~\bibnamefont {Zhu}},\ }in\ \href
  {\doibase 10.1007/978-3-540-73545-8_21} {\emph {\bibinfo {booktitle}
  {Computing and Combinatorics}}},\ \bibinfo {series} {Lecture Notes in
  Computer Science}, Vol.\ \bibinfo {volume} {4598},\ \bibinfo {editor} {edited
  by\ \bibinfo {editor} {\bibfnamefont {G.}~\bibnamefont {Lin}}}\ (\bibinfo
  {year} {2007})\ pp.\ \bibinfo {pages} {198--209}\BibitemShut {NoStop}%
\bibitem [{\citenamefont {Ashwin}\ \emph {et~al.}(2012)\citenamefont {Ashwin},
  \citenamefont {Blawzdziewicz}, \citenamefont {O'Hern},\ and\ \citenamefont
  {Shattuck}}]{Ashwin12}%
  \BibitemOpen
  \bibfield  {author} {\bibinfo {author} {\bibfnamefont {S.~S.}\ \bibnamefont
  {Ashwin}}, \bibinfo {author} {\bibfnamefont {J.}~\bibnamefont
  {Blawzdziewicz}}, \bibinfo {author} {\bibfnamefont {C.~S.}\ \bibnamefont
  {O'Hern}}, \ and\ \bibinfo {author} {\bibfnamefont {M.~D.}\ \bibnamefont
  {Shattuck}},\ }\href {\doibase 10.1103/PhysRevE.85.061307} {\bibfield
  {journal} {\bibinfo  {journal} {Physical Review E}\ }\textbf {\bibinfo
  {volume} {85}},\ \bibinfo {pages} {061307} (\bibinfo {year}
  {2012})}\BibitemShut {NoStop}%
\bibitem [{\citenamefont {Xu}\ \emph {et~al.}(2011)\citenamefont {Xu},
  \citenamefont {Frenkel},\ and\ \citenamefont {Liu}}]{Xu11}%
  \BibitemOpen
  \bibfield  {author} {\bibinfo {author} {\bibfnamefont {N.}~\bibnamefont
  {Xu}}, \bibinfo {author} {\bibfnamefont {D.}~\bibnamefont {Frenkel}}, \ and\
  \bibinfo {author} {\bibfnamefont {A.~J.}\ \bibnamefont {Liu}},\ }\href
  {\doibase 10.1103/PhysRevLett.106.245502} {\bibfield  {journal} {\bibinfo
  {journal} {Physical Review Letters}\ }\textbf {\bibinfo {volume} {106}},\
  \bibinfo {pages} {245502} (\bibinfo {year} {2011})}\BibitemShut {NoStop}%
\bibitem [{\citenamefont {Asenjo}\ \emph {et~al.}(2013)\citenamefont {Asenjo},
  \citenamefont {Stevenson}, \citenamefont {Wales},\ and\ \citenamefont
  {Frenkel}}]{Asenjo13}%
  \BibitemOpen
  \bibfield  {author} {\bibinfo {author} {\bibfnamefont {D.}~\bibnamefont
  {Asenjo}}, \bibinfo {author} {\bibfnamefont {J.~D.}\ \bibnamefont
  {Stevenson}}, \bibinfo {author} {\bibfnamefont {D.~J.}\ \bibnamefont
  {Wales}}, \ and\ \bibinfo {author} {\bibfnamefont {D.}~\bibnamefont
  {Frenkel}},\ }\href {\doibase 10.1021/jp312457a} {\bibfield  {journal}
  {\bibinfo  {journal} {Journal of Physical Chemistry B}\ }\textbf {\bibinfo
  {volume} {117}},\ \bibinfo {pages} {12717} (\bibinfo {year}
  {2013})}\BibitemShut {NoStop}%
\bibitem [{\citenamefont {Asenjo}\ \emph {et~al.}(2014)\citenamefont {Asenjo},
  \citenamefont {Paillusson},\ and\ \citenamefont {Frenkel}}]{Asenjo14}%
  \BibitemOpen
  \bibfield  {author} {\bibinfo {author} {\bibfnamefont {D.}~\bibnamefont
  {Asenjo}}, \bibinfo {author} {\bibfnamefont {F.}~\bibnamefont {Paillusson}},
  \ and\ \bibinfo {author} {\bibfnamefont {D.}~\bibnamefont {Frenkel}},\ }\href
  {\doibase 10.1103/physrevlett.112.098002} {\bibfield  {journal} {\bibinfo
  {journal} {Physical Review Letters}\ }\textbf {\bibinfo {volume} {112}},\
  \bibinfo {pages} {098002} (\bibinfo {year} {2014})}\BibitemShut {NoStop}%
\bibitem [{\citenamefont {Martiniani}\ \emph {et~al.}(2016)\citenamefont
  {Martiniani}, \citenamefont {Schrenk}, \citenamefont {Stevenson},
  \citenamefont {Wales},\ and\ \citenamefont {Frenkel}}]{Martiniani16}%
  \BibitemOpen
  \bibfield  {author} {\bibinfo {author} {\bibfnamefont {S.}~\bibnamefont
  {Martiniani}}, \bibinfo {author} {\bibfnamefont {K.~J.}\ \bibnamefont
  {Schrenk}}, \bibinfo {author} {\bibfnamefont {J.~D.}\ \bibnamefont
  {Stevenson}}, \bibinfo {author} {\bibfnamefont {D.~J.}\ \bibnamefont
  {Wales}}, \ and\ \bibinfo {author} {\bibfnamefont {D.}~\bibnamefont
  {Frenkel}},\ }\href {\doibase 10.1103/PhysRevE.93.012906} {\bibfield
  {journal} {\bibinfo  {journal} {Phys. Rev. E}\ }\textbf {\bibinfo {volume}
  {93}},\ \bibinfo {pages} {012906} (\bibinfo {year} {2016})}\BibitemShut
  {NoStop}%
\bibitem [{\citenamefont {Xu}\ \emph {et~al.}(2010)\citenamefont {Xu},
  \citenamefont {Vitelli}, \citenamefont {Liu},\ and\ \citenamefont
  {Nagel}}]{Xu10}%
  \BibitemOpen
  \bibfield  {author} {\bibinfo {author} {\bibfnamefont {N.}~\bibnamefont
  {Xu}}, \bibinfo {author} {\bibfnamefont {V.}~\bibnamefont {Vitelli}},
  \bibinfo {author} {\bibfnamefont {A.~J.}\ \bibnamefont {Liu}}, \ and\
  \bibinfo {author} {\bibfnamefont {S.~R.}\ \bibnamefont {Nagel}},\ }\href
  {\doibase 10.1209/0295-5075/90/56001} {\bibfield  {journal} {\bibinfo
  {journal} {EPL}\ }\textbf {\bibinfo {volume} {90}},\ \bibinfo {pages} {56001}
  (\bibinfo {year} {2010})}\BibitemShut {NoStop}%
\bibitem [{\citenamefont {Wang}\ \emph {et~al.}(2012)\citenamefont {Wang},
  \citenamefont {Song}, \citenamefont {Wang},\ and\ \citenamefont
  {Makse}}]{Wang12}%
  \BibitemOpen
  \bibfield  {author} {\bibinfo {author} {\bibfnamefont {K.}~\bibnamefont
  {Wang}}, \bibinfo {author} {\bibfnamefont {C.}~\bibnamefont {Song}}, \bibinfo
  {author} {\bibfnamefont {P.}~\bibnamefont {Wang}}, \ and\ \bibinfo {author}
  {\bibfnamefont {H.~A.}\ \bibnamefont {Makse}},\ }\href {\doibase
  10.1103/PhysRevE.86.011305} {\bibfield  {journal} {\bibinfo  {journal}
  {Physical Review E}\ }\textbf {\bibinfo {volume} {86}},\ \bibinfo {pages}
  {011305} (\bibinfo {year} {2012})}\BibitemShut {NoStop}%
\bibitem [{\citenamefont {Milman}(1998)}]{milman1998surprising}%
  \BibitemOpen
  \bibfield  {author} {\bibinfo {author} {\bibfnamefont {V.}~\bibnamefont
  {Milman}},\ }in\ \href {\doibase 10.1007/978-3-0348-8898-1_4} {\emph
  {\bibinfo {booktitle} {European Congress of Mathematics}}}\ (\bibinfo
  {publisher} {Birkhäuser},\ \bibinfo {address} {Basel},\ \bibinfo {year}
  {1998})\ pp.\ \bibinfo {pages} {73--91}\BibitemShut {NoStop}%
\bibitem [{\citenamefont {Guruswami}\ and\ \citenamefont
  {Kannan}(2012)}]{Guruswami2012}%
  \BibitemOpen
  \bibfield  {author} {\bibinfo {author} {\bibfnamefont {V.}~\bibnamefont
  {Guruswami}}\ and\ \bibinfo {author} {\bibfnamefont {R.}~\bibnamefont
  {Kannan}},\ }\href
  {https://www.cs.cmu.edu/~venkatg/teaching/CStheory-infoage/} {\enquote
  {\bibinfo {title} {Computer science theory for the information age},}\ }
  (\bibinfo {year} {2012}),\ \bibinfo {note}
  {\url{https://www.cs.cmu.edu/~venkatg/teaching/CStheory-infoage/}}\BibitemShut
  {NoStop}%
\bibitem [{\citenamefont {Steinhardt}\ \emph {et~al.}(1983)\citenamefont
  {Steinhardt}, \citenamefont {Nelson},\ and\ \citenamefont
  {Ronchetti}}]{steinhardt1983bond}%
  \BibitemOpen
  \bibfield  {author} {\bibinfo {author} {\bibfnamefont {P.~J.}\ \bibnamefont
  {Steinhardt}}, \bibinfo {author} {\bibfnamefont {D.~R.}\ \bibnamefont
  {Nelson}}, \ and\ \bibinfo {author} {\bibfnamefont {M.}~\bibnamefont
  {Ronchetti}},\ }\href {\doibase http://dx.doi.org/10.1103/PhysRevB.28.784}
  {\bibfield  {journal} {\bibinfo  {journal} {Physical Review B}\ }\textbf
  {\bibinfo {volume} {28}},\ \bibinfo {pages} {784} (\bibinfo {year}
  {1983})}\BibitemShut {NoStop}%
\bibitem [{\citenamefont {O'Hern}\ \emph {et~al.}(2003)\citenamefont {O'Hern},
  \citenamefont {Silbert}, \citenamefont {Liu},\ and\ \citenamefont
  {Nagel}}]{OHern03}%
  \BibitemOpen
  \bibfield  {author} {\bibinfo {author} {\bibfnamefont {C.~S.}\ \bibnamefont
  {O'Hern}}, \bibinfo {author} {\bibfnamefont {L.~E.}\ \bibnamefont {Silbert}},
  \bibinfo {author} {\bibfnamefont {A.~J.}\ \bibnamefont {Liu}}, \ and\
  \bibinfo {author} {\bibfnamefont {S.~R.}\ \bibnamefont {Nagel}},\ }\href
  {\doibase 10.1103/PhysRevE.68.011306} {\bibfield  {journal} {\bibinfo
  {journal} {Physical Review E}\ }\textbf {\bibinfo {volume} {68}},\ \bibinfo
  {pages} {011306} (\bibinfo {year} {2003})}\BibitemShut {NoStop}%
\bibitem [{\citenamefont {van Meel}\ \emph {et~al.}(2012)\citenamefont {van
  Meel}, \citenamefont {Filion}, \citenamefont {Valeriani},\ and\ \citenamefont
  {Frenkel}}]{vanMeel12}%
  \BibitemOpen
  \bibfield  {author} {\bibinfo {author} {\bibfnamefont {J.}~\bibnamefont {van
  Meel}}, \bibinfo {author} {\bibfnamefont {L.}~\bibnamefont {Filion}},
  \bibinfo {author} {\bibfnamefont {C.}~\bibnamefont {Valeriani}}, \ and\
  \bibinfo {author} {\bibfnamefont {D.}~\bibnamefont {Frenkel}},\ }\href
  {http://scitation.aip.org/content/aip/journal/jcp/136/23/10.1063/1.4729313}
  {\bibfield  {journal} {\bibinfo  {journal} {The Journal of Chemical Physics}\
  }\textbf {\bibinfo {volume} {136}},\ \bibinfo {eid} {234107} (\bibinfo {year}
  {2012})}\BibitemShut {NoStop}%
\bibitem [{\citenamefont {Rudnick}\ and\ \citenamefont
  {Gaspari}(1987)}]{rudnick1987shapes}%
  \BibitemOpen
  \bibfield  {author} {\bibinfo {author} {\bibfnamefont {J.}~\bibnamefont
  {Rudnick}}\ and\ \bibinfo {author} {\bibfnamefont {G.}~\bibnamefont
  {Gaspari}},\ }\href {\doibase 10.1126/science.237.4813.384} {\bibfield
  {journal} {\bibinfo  {journal} {Science}\ }\textbf {\bibinfo {volume}
  {237}},\ \bibinfo {pages} {384} (\bibinfo {year} {1987})}\BibitemShut
  {NoStop}%
\end{thebibliography}%


\begin{thebibliography}{10}%
\makeatletter
\providecommand \@ifxundefined [1]{%
 \@ifx{#1\undefined}
}%
\providecommand \@ifnum [1]{%
 \ifnum #1\expandafter \@firstoftwo
 \else \expandafter \@secondoftwo
 \fi
}%
\providecommand \@ifx [1]{%
 \ifx #1\expandafter \@firstoftwo
 \else \expandafter \@secondoftwo
 \fi
}%
\providecommand \natexlab [1]{#1}%
\providecommand \enquote  [1]{``#1''}%
\providecommand \bibnamefont  [1]{#1}%
\providecommand \bibfnamefont [1]{#1}%
\providecommand \citenamefont [1]{#1}%
\providecommand \href@noop [0]{\@secondoftwo}%
\providecommand \href [0]{\begingroup \@sanitize@url \@href}%
\providecommand \@href[1]{\@@startlink{#1}\@@href}%
\providecommand \@@href[1]{\endgroup#1\@@endlink}%
\providecommand \@sanitize@url [0]{\catcode `\\12\catcode `\$12\catcode
  `\&12\catcode `\#12\catcode `\^12\catcode `\_12\catcode `\%12\relax}%
\providecommand \@@startlink[1]{}%
\providecommand \@@endlink[0]{}%
\providecommand \url  [0]{\begingroup\@sanitize@url \@url }%
\providecommand \@url [1]{\endgroup\@href {#1}{\urlprefix }}%
\providecommand \urlprefix  [0]{URL }%
\providecommand \Eprint [0]{\href }%
\providecommand \doibase [0]{http://dx.doi.org/}%
\providecommand \selectlanguage [0]{\@gobble}%
\providecommand \bibinfo  [0]{\@secondoftwo}%
\providecommand \bibfield  [0]{\@secondoftwo}%
\providecommand \translation [1]{[#1]}%
\providecommand \BibitemOpen [0]{}%
\providecommand \bibitemStop [0]{}%
\providecommand \bibitemNoStop [0]{.\EOS\space}%
\providecommand \EOS [0]{\spacefactor3000\relax}%
\providecommand \BibitemShut  [1]{\csname bibitem#1\endcsname}%
\let\auto@bib@innerbib\@empty
\bibitem [{\citenamefont {Martiniani}\ \emph {et~al.}(2016)\citenamefont
  {Martiniani}, \citenamefont {Schrenk}, \citenamefont {Stevenson},
  \citenamefont {Wales},\ and\ \citenamefont {Frenkel}}]{Martiniani16}%
  \BibitemOpen
  \bibfield  {author} {\bibinfo {author} {\bibfnamefont {Stefano}\ \bibnamefont
  {Martiniani}}, \bibinfo {author} {\bibfnamefont {K.~Julian}\ \bibnamefont
  {Schrenk}}, \bibinfo {author} {\bibfnamefont {Jacob~D.}\ \bibnamefont
  {Stevenson}}, \bibinfo {author} {\bibfnamefont {David~J.}\ \bibnamefont
  {Wales}}, \ and\ \bibinfo {author} {\bibfnamefont {Daan}\ \bibnamefont
  {Frenkel}},\ }\bibfield  {title} {\enquote {\bibinfo {title} {Turning
  intractable counting into sampling: Computing the configurational entropy of
  three-dimensional jammed packings},}\ }\href {\doibase
  10.1103/PhysRevE.93.012906} {\bibfield  {journal} {\bibinfo  {journal} {Phys.
  Rev. E}\ }\textbf {\bibinfo {volume} {93}},\ \bibinfo {pages} {012906}
  (\bibinfo {year} {2016})}\BibitemShut {NoStop}%
\bibitem [{Note1()}]{Note1}%
  \BibitemOpen
  \bibinfo {note} {Note that for convenience we used the same choice of
  positive $k$'s as required for thermodynamic integration. For this particular
  method any choice of $k$'s is appropriate, typically a geometric
  distribution, denser for small $k$ and coarser near $k_{\protect \text {1}}$
  is also suitable.}\BibitemShut {Stop}%
\bibitem [{Note2()}]{Note2}%
  \BibitemOpen
  \bibinfo {note} {To do so we sample a direction from the surface of the unit
  sphere and the length of the displacement from a $\protect \text
  {Normal}(0,\sigma )$.}\BibitemShut {Stop}%
\bibitem [{\citenamefont {Shirts}\ and\ \citenamefont
  {Chodera}(2008)}]{Shirts2008}%
  \BibitemOpen
  \bibfield  {author} {\bibinfo {author} {\bibfnamefont {M.~R.}\ \bibnamefont
  {Shirts}}\ and\ \bibinfo {author} {\bibfnamefont {J.~D.}\ \bibnamefont
  {Chodera}},\ }\bibfield  {title} {\enquote {\bibinfo {title} {Statistically
  optimal analysis of samples from multiple equilibrium states},}\ }\href
  {\doibase 10.1063/1.2978177} {\bibfield  {journal} {\bibinfo  {journal}
  {Journal of Chemical Physics}\ }\textbf {\bibinfo {volume} {129}},\ \bibinfo
  {eid} {124105} (\bibinfo {year} {2008})}\BibitemShut {NoStop}%
\bibitem [{PyM()}]{PyMBAR}%
  \BibitemOpen
  \href {https://github.com/choderalab/pymbar} {}\bibinfo {note}
  {\url{https://github.com/choderalab/pymbar}}\BibitemShut {NoStop}%
\bibitem [{\citenamefont {Bishop}(2009)}]{Bishop09}%
  \BibitemOpen
  \bibfield  {author} {\bibinfo {author} {\bibfnamefont {C.~M.}\ \bibnamefont
  {Bishop}},\ }\href@noop {} {\emph {\bibinfo {title} {Pattern recognition and
  machine learning}}}\ (\bibinfo  {publisher} {Springer},\ \bibinfo {address}
  {New York},\ \bibinfo {year} {2009})\BibitemShut {NoStop}%
\bibitem [{\citenamefont {Weeks}\ \emph {et~al.}(1971)\citenamefont {Weeks},
  \citenamefont {Chandler},\ and\ \citenamefont {Andersen}}]{Weeks71}%
  \BibitemOpen
  \bibfield  {author} {\bibinfo {author} {\bibfnamefont {John~D}\ \bibnamefont
  {Weeks}}, \bibinfo {author} {\bibfnamefont {David}\ \bibnamefont {Chandler}},
  \ and\ \bibinfo {author} {\bibfnamefont {Hans~C}\ \bibnamefont {Andersen}},\
  }\bibfield  {title} {\enquote {\bibinfo {title} {Role of repulsive forces in
  determining the equilibrium structure of simple liquids},}\ }\href {\doibase
  10.1063/1.1674820} {\bibfield  {journal} {\bibinfo  {journal} {Journal of
  Chemical Physics}\ }\textbf {\bibinfo {volume} {54}},\ \bibinfo {pages}
  {5237} (\bibinfo {year} {1971})}\BibitemShut {NoStop}%
\bibitem [{\citenamefont {Hager}\ and\ \citenamefont {Zhang}(2005)}]{Hager05}%
  \BibitemOpen
  \bibfield  {author} {\bibinfo {author} {\bibfnamefont {William~W}\
  \bibnamefont {Hager}}\ and\ \bibinfo {author} {\bibfnamefont {Hongchao}\
  \bibnamefont {Zhang}},\ }\bibfield  {title} {\enquote {\bibinfo {title} {A
  new conjugate gradient method with guaranteed descent and an efficient line
  search},}\ }\href {\doibase 10.1137/030601880} {\bibfield  {journal}
  {\bibinfo  {journal} {SIAM Journal on Optimization}\ }\textbf {\bibinfo
  {volume} {16}},\ \bibinfo {pages} {170} (\bibinfo {year} {2005})}\BibitemShut
  {NoStop}%
\bibitem [{\citenamefont {Hager}\ and\ \citenamefont {Zhang}(2006)}]{Hager06}%
  \BibitemOpen
  \bibfield  {author} {\bibinfo {author} {\bibfnamefont {William~W}\
  \bibnamefont {Hager}}\ and\ \bibinfo {author} {\bibfnamefont {Hongchao}\
  \bibnamefont {Zhang}},\ }\bibfield  {title} {\enquote {\bibinfo {title}
  {Algorithm 851: {CG}\_{DESCENT}, a conjugate gradient method with guaranteed
  descent},}\ }\href {\doibase 10.1145/1132973.1132979} {\bibfield  {journal}
  {\bibinfo  {journal} {ACM Transactions on Mathematical Software}\ }\textbf
  {\bibinfo {volume} {32}},\ \bibinfo {pages} {113} (\bibinfo {year}
  {2006})}\BibitemShut {NoStop}%
\bibitem [{PyC()}]{PyCG_DESCENT}%
  \BibitemOpen
  \href {https://github.com/smcantab/PyCG_DESCENT} {}\bibinfo {note}
  {\url{https://github.com/smcantab/PyCG_DESCENT}}\BibitemShut {NoStop}%
\end{thebibliography}%
\end{document}


\title{Supplemental material: Structural analysis of high-dimensional
  basins of attraction}
\author{Stefano Martiniani}
\email{sm958@cam.ac.uk}
\affiliation{Department of Chemistry, University of Cambridge,
  Lensfield Road, Cambridge, CB2 1EW, UK}
\author{K. Julian Schrenk}
\affiliation{Department of Chemistry, University of Cambridge,
  Lensfield Road, Cambridge, CB2 1EW, UK}
\author{Jacob D. Stevenson}
\affiliation{Microsoft Research Ltd, 21 Station Road, Cambridge, CB1
  2FB, UK}
\affiliation{Department of Chemistry, University of Cambridge, 
Lensfield Road, Cambridge, CB2 1EW, UK}
\author{David J. Wales}
\affiliation{Department of Chemistry, University of Cambridge,
  Lensfield Road, Cambridge, CB2 1EW, UK}
\author{Daan Frenkel}
\affiliation{Department of Chemistry, University of Cambridge,
  Lensfield Road, Cambridge, CB2 1EW, UK}
\maketitle
\section{\label{si::sec::volume_computation} Volume computation}
\begin{figure}
    \includegraphics[width=0.5\columnwidth]{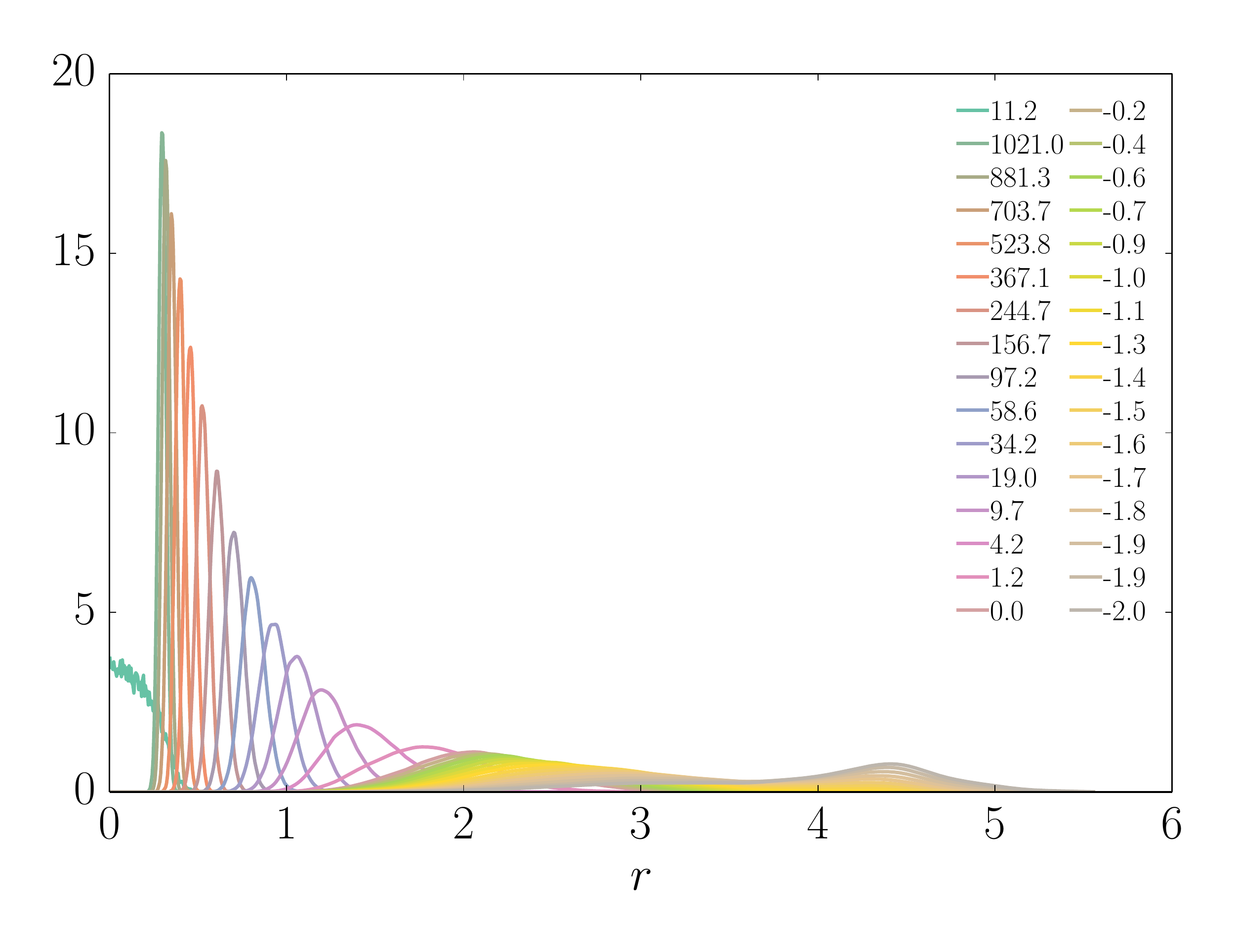}
    \caption{\label{si::fig::histogram} Kernel density estimation of
      the distance sampled by a random walk within the basin, coupled
      to the minimum with decreasing coupling constant from left to
      right. The left-most curve was obtained by direct sampling as
      described in the main text. Replicas with negative coupling
      constants explore regions of the volume that would otherwise
      never be visited. This particular example is of a disordered
      packing with polidispersity $\eta=0.037$.}
\end{figure}

We choose a set of harmonic bias potential functions
\begin{equation} 
u_i = \frac{1}{2}k_i|\vect{x}-\vect{x}_0|^2 
\label{eq:ui}
\end{equation}
with $k_{i \in [1,M]}=\{k_{1},\dots,k_{M-1},0\}$ and perform $10^{10}$
Hamiltonian Parallel Tempering steps as described in
Ref.~\cite{Martiniani16} \footnote{Note that for convenience we used
  the same choice of positive $k$'s as required for thermodynamic
  integration. For this particular method any choice of $k$'s is
  appropriate, typically a geometric distribution, denser for small
  $k$ and coarser near $k_{\text{1}}$ is also suitable.}. Note that
each Monte Carlo step is followed by a full energy minimisation to
test whether the walker has stepped outside the basin of
attraction. We choose half of the $k$'s to be positive and the other
half negative to accelerate equilibration as well as to increase the
DOS resolution near the boundary of the basin. The distributions
obtained from the replicas with negative coupling constants contribute
to the final MBAR volume estimation, unlike for TI. We stress that the
choice of biasing potential is arbitrary.  Near the origin we sample
the set of configurations $\{\vect{x}\}_0$ directly from a hypersphere
centred at $\vect{x}_0$ with radius sampled from a Gaussian
distribution with standard deviation
$\sigma=\sqrt{\langle|\vect{x}-\vect{x}_0|^2\rangle_{k_1}}$,
corresponding to a coupling constant $k_0 = 1/\sigma^2$. This choice
of $k_0$ is such that there is sufficient overlap between the
distributions of $\{\vect{x}\}_0$ and $\{\vect{x}\}_1$, as can be
verified looking at the two leftmost curves in
Fig.~(\ref{si::fig::histogram}) \footnote{To do so we sample a
  direction from the surface of the unit sphere and the length of the
  displacement from a $\text{Normal}(0,\sigma)$.}. The corresponding
bias potential function is
\begin{equation}
u_0 =
(n-1)\log|\vect{x}-\vect{x}_0|+\frac{1}{2}k_0|\vect{x}-\vect{x}_0|^2,
\label{eq:u0}
\end{equation}
where the first term on the right-hand-side is the $\log$-DOS for a
$n$-ball, necessary to account for the greater entropy associated with
the regions of space further away from the origin. For a system of $N$
particles in $d$ dimensions with fixed centre of mass we have
$n=(N-1)d$ degrees of freedom. The overhead associated with this
calculation is insignificant compared to the Hamiltonian Parallel
Tempering since the samples thus drawn are completely uncorrelated.

We compute the reduced free energy differences between each of $1+31$
replicas with reduced potential functions given by
Eqs.~(\ref{eq:ui})--(\ref{eq:u0}) using PyMBAR \cite{Shirts2008,
  PyMBAR}. As reference volume we choose $\omega$ to be the $n$-ball
of radius $r_{\omega}$ centred on $\vect{x}_0$ with approximately
$\mathcal{R}=0.9$ of its volume contained within the basin
$\Omega$. We choose $\omega \not\subseteq \Omega$ to allow more
samples with $|\vect{x}-\vect{x}_0| \leq r_{\omega}$ thus reducing the
uncertainty in the MBAR estimate. For $\mathcal{R} \approx 1$ we can
correct exactly for this by noting that
\begin{equation}
\mathcal{R} =
\frac{1}{V_{\omega}}\int_{\omega} p_0(\vect{x})d\vect{x}
\end{equation} 
and $\mathcal{R}$ can be computed directly by Monte Carlo.  We thus
rewrite Eq.~(9) as
\begin{equation}
f_{M} = f_{\omega} - \log\mathcal{R} + (\hat{f}_{M}
- \hat{f}_{\omega}).
\end{equation} 
Note that the difference in reduced free energies computed using a
reference sphere of radius $r_{\omega}/2$ or $2r_{\omega}$ is within
the statistical uncertainty, hence the method is robust with respect
to the choice of reference sphere. We also note that this method ought
not be limited to the $n$-ball as the choice of reference volume, in
fact any geometrical body $\omega \subseteq \Omega$ of known volume
and surface (thus for which a similar expression to Eq.~(\ref{eq:u0})
can be derived) is suitable, for instance a hypercube or a
hyperellipsoid. If $\omega \not\subseteq \Omega$ then an accurate
estimate of $\mathcal{R}$ must be available.

\subsection{\label{si:sec:dos} Density of states}

From the analysis of the posterior probability density functions, the
present method may yield structural information, as an easy to compute
by-product. Choosing a set of biasing potentials $u_i(r)$ that are a
function of the distance from the origin $r=|\vect{x}-\vect{x}_0|$, we
can compute the overall density of states (DOS) for the manifold as a
function of $r$. From each of the $K+1$ replicas' trajectories
$\{\vect{x}\}_i$ we obtain a (binless) kernel density estimation (KDE)
\cite{Bishop09} of the probability density functions $h_i(r)$, see
Fig.~(\ref{si::fig::histogram}) for an example, which must be unbiased
and summed over all replicas to obtain the overall log-DOS function as
\begin{equation}
\log h(r) = \sum_{i=0}^{K} w_i(r) \left(\log h_i(r) + u_i(r) -
\Delta\hat{f}_{0i} \right) .
\label{eq:logdos}
\end{equation}
where $w_i(r) = h_i(r)/\sum_{i=0}^{K} h_i(r)$ are normalised weights
and $\Delta\hat{f}_{0i}$ are the free energy differences between
replicas $R_i$ and $R_0$.

\subsection{\label{si:sec:ticomp} Comparison to thermodynamic integration}

\begin{figure}
    \includegraphics[width=0.5\columnwidth]{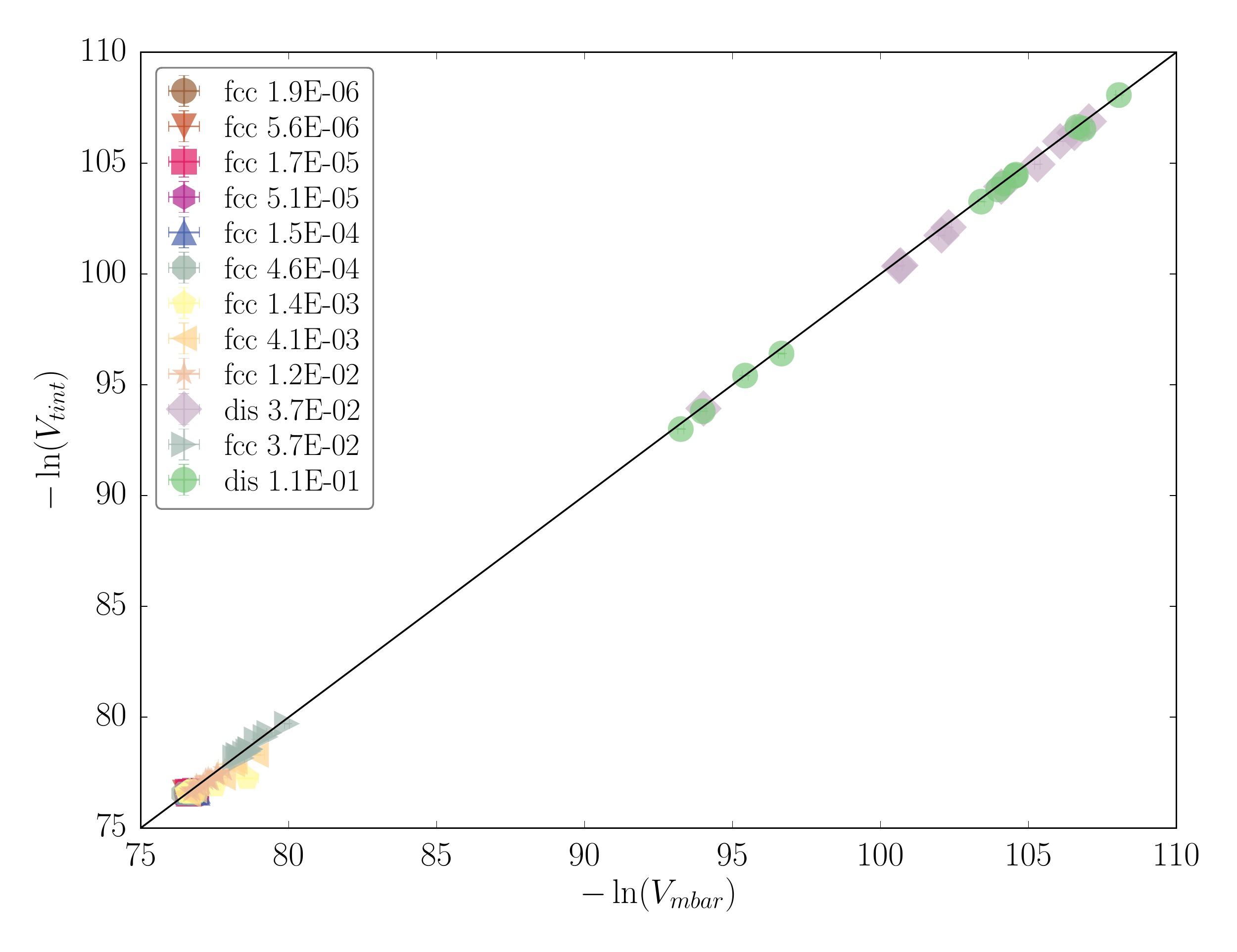}
    \caption{\label{si::fig::mbar_tint_comparison} Comparison of the
      volumes computed by thermodynamic integration, using only the
      replicas with positive coupling constant, and by MBAR following
      the protocol described in this work.}
\end{figure}

\begin{figure}
    \includegraphics[width=0.5\columnwidth]{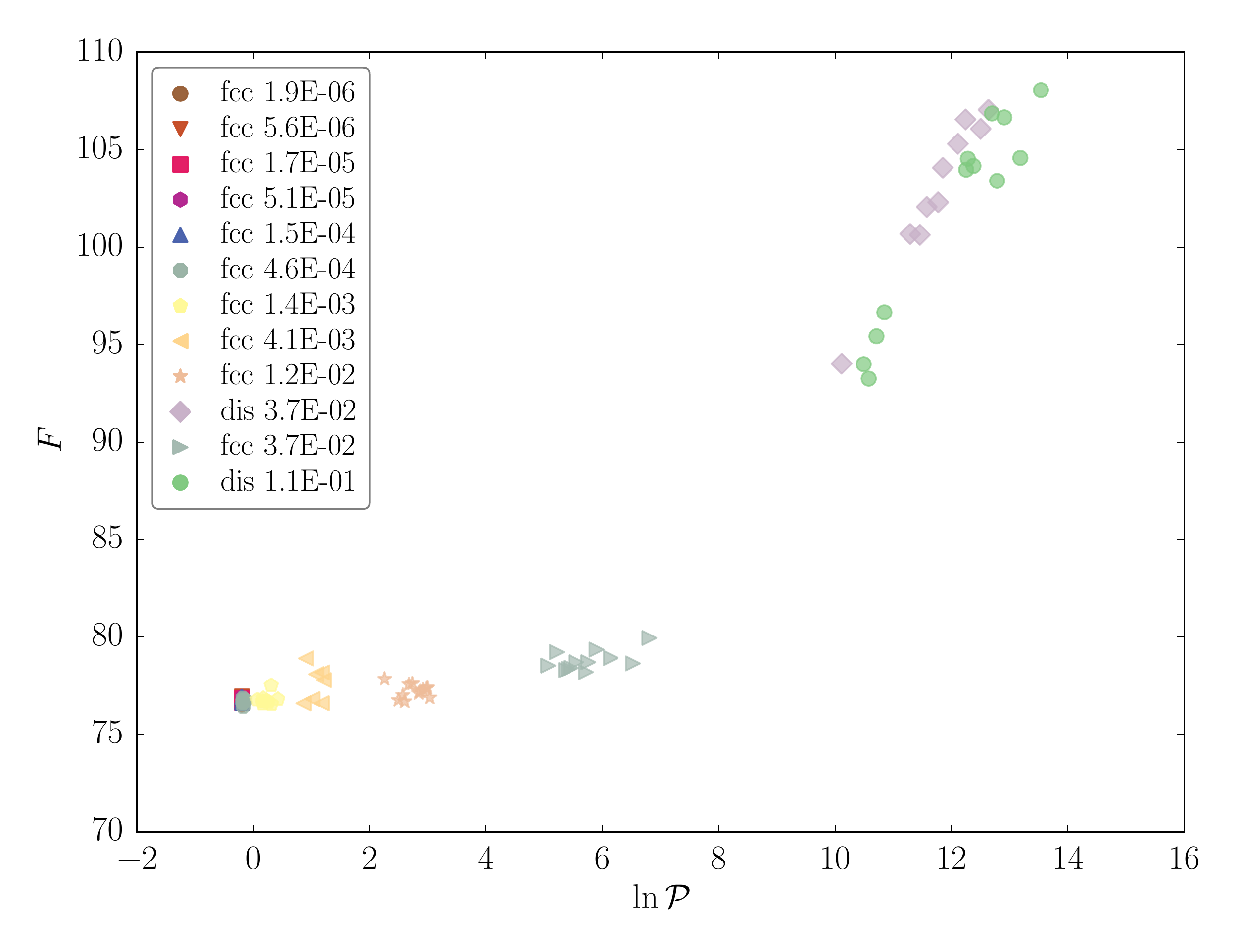}
    \caption{\label{si::fig::f_logp} Dimensionless free energy
      ($F \equiv -\ln V$) versus pressure of the mechanically stable
      states analysed in the main text. For details of the the strong
      correlation between pressure and volume refer to Martiniani et
      al.\cite{Martiniani16}.}
\end{figure}

\newpage

\section{\label{si:sec:ticomp} Jammed packings of polydisperse HS-WCA spheres}

We draw $N=32$ particle radii $\{r_{\text{HS}}\}_N$ from a Gaussian
distribution $\mathrm{Normal}(1,\eta) > 0$, truncated at
$r_{\text{HS}}=0$, set the box size to meet the target packing
fraction of the hard sphere fluid $\phi_\text{HS}$ and then place the
particles in a valid, either fcc or random, initial hard sphere
configuration, as described in the main text.

Given these hard sphere configurations, we switch on a soft repulsive
interaction to generate over-compressed jammed packings of the
particles and relax the system to a mechanically stable state by
energy minimization. The particles are inflated with a WCA-like
potential \cite{Weeks71} to reach the target soft packing fraction
$\phi_\text{SS} > \phi_\text{HS}^{(\text{RCP})} > \phi_\text{HS}$.
The hard spheres are inflated proportional to their radius, so that
the soft sphere radius is
\begin{equation}
    r_{\text{SS}}
    =\left(\frac{\phi_\text{SS}}{\phi_\text{HS}}\right)^{1/d}
    r_{\text{HS}},
\end{equation}
where $d$ is the dimensionality of the box, $r_{\text{SS}}$ and
$r_{\text{HS}}$ the soft and hard sphere radii respectively. Clearly,
this procedure does not change the polydispersity of the sample.

We define the WCA-like potential around a hard core as follows:
consider two spherical particles with hard core distance
$r_{\text{HS}}$ and soft core contact distance
$r_{\text{SS}}=r_{\text{HS}}(1+\theta)$, with $\theta =
(\phi_\text{SS}/\phi_\text{HS})^{1/d}-1$. We can then write a
horizontally shifted hard-sphere plus WCA (HS-WCA) potential as
\begin{equation}
  \label{eq:hswca}
    v_\text{HS-WCA}(r) = 
    \left\{ 
    \begin{array}{l l}
        \infty &\quad r \leq r_{\text{HS}},\\
        \begin{array}{l}
        4\epsilon \left[
          \left(\frac{\displaystyle\sigma(r_{\text{HS}})}{\displaystyle
            r^2-r_{\text{HS}}^2} \right)^{12}
          \right. \\ \left. -\left(\frac{\displaystyle\sigma(r_{\text{HS}})}{\displaystyle
            r^2-r_{\text{HS}}^2} \right)^6 \right] + \epsilon
        \end{array}
        &\quad r_{\text{HS}} < r < r_{\text{SS}}, \\ 0 &\quad r \geq
        r_{\text{SS}}
        \end{array}
    \right.
\end{equation}
where $\sigma(r_{\text{HS}}) = (2\theta +\theta^2)
r_{\text{HS}}^2/2^{1/6}$ guarantees that the potential goes to zero at
$r_{\text{SS}}$. For computational convenience (avoidance of
square-root evaluations), the potential in Eq.~\ref{eq:hswca} differs
from the WCA form in that the inter-particle distance in the
denominator of the WCA potential has been replaced with a difference
of squares.

Numerically evaluating this potential, we match the gradient and
linearly continue the function $v_\text{HS-WCA}(r)$ for $r \leq
r_{\text{HS}} + \varepsilon$, with $\varepsilon > 0$ an arbitrary
small constant, such that minimisation is still meaningful if hard
core overlaps do occur.

Energy minimisations are performed with the CG$\_$DESCENT algorithm
\cite{Hager05, Hager06, PyCG_DESCENT}.

\bibliography{mbar_bv_method_si}